\documentclass[twocolumn,showkeys,aps,prb,showpacs]{revtex4-1}
\UseRawInputEncoding
\usepackage{graphicx}
\usepackage[CJKbookmarks,dvipdfm,colorlinks,linkcolor=red,citecolor=blue]{hyperref}

\begin{document}

\title{Valley-polarized quantum anomalous Hall insulator in monolayer $\mathrm{RuBr_2}$}

\author{San-Dong Guo$^{1}$, Wen-Qi Mu$^{1}$ and  Bang-Gui Liu$^{2,3}$}
\affiliation{$^1$School of Electronic Engineering, Xi'an University of Posts and Telecommunications, Xi'an 710121, China}
\affiliation{$^2$ Beijing National Laboratory for Condensed Matter Physics, Institute of Physics, Chinese Academy of Sciences, Beijing 100190, People's Republic of China}
\affiliation{$^3$School of Physical Sciences, University of Chinese Academy of Sciences, Beijing 100190, People's Republic of China}
\begin{abstract}
Coexistence of intrinsic ferrovalley (FV) and nontrivial band topology  attracts intensive interest both for its fundamental
physics and for its potential applications,  namely valley-polarized quantum anomalous Hall insulator (VQAHI).
Here, based on first-principles calculations by using  generalized gradient
approximation plus $U$ (GGA+$U$) approach,  the VQAHI induced by electronic correlation or strain can occur
in  monolayer $\mathrm{RuBr_2}$.
For perpendicular magnetic anisotropy (PMA), the ferrovalley (FV) to half-valley-metal (HVM)  to quantum anomalous Hall (QAH) to HVM to FV transitions can be driven by increasing  electron correlation $U$. However, there are  no special QAH states and  valley polarization for in-plane magnetic anisotropy.
By calculating actual magnetic anisotropy energy (MAE), the VQAHI indeed can exist between two HVM states due to PMA, a unit Chern number/a chiral edge state and spontaneous valley polarization. The increasing $U$ can induce VQAHI, which can be explained by sign-reversible  Berry curvature or   band inversion  between $d_{xy}$/$d_{x^2-y^2}$ and $d_{z^2}$ orbitals. Even though the real $U$ falls outside the range, the VQAHI can be achieved by strain.
Taking $U$$=$2.25 eV as a concrete case, the monolayer $\mathrm{RuBr_2}$ can change from a common ferromagentic (FM) semiconductor to VQAHI under about  0.985 compressive strain.  It is noted that  the edge states of VQAHI are
chiral-spin-valley locking, which can achieve  complete spin and valley polarizations for  low-dissipation electronics devices.
Both energy band gap and valley splitting of VQAHI in monolayer $\mathrm{RuBr_2}$ are higher than
the thermal energy of  room temperature (25 meV), which is key at room temperature for
device applications. It is found that electronic correlation or strain have important effects on Curie temperature of monolayer $\mathrm{RuBr_2}$.
These results  can be readily extended to other monolayer  $\mathrm{MXY}$ (M = Ru, Os; X/Y=Cl, Br I).
Our works emphasize  the importance of electronic correlation and PMA to study FV materials, and provide a pathway to realize VQAHI.

\end{abstract}
\keywords{Ferrovalley, Band topology, Electronic correlation, Strain~~~~~~~~~~~~~~~~~~~~~~~~~~~~~~~~~~Email:sandongyuwang@163.com}

\maketitle

\section{Introduction}
The valleys,  namely the extrema on the valence/conductuion band, provide a different degree of freedom to  manipulate information analogous to charge
and spin (called as  valleytronics), which has
attracted intensive attention\cite{q1,q2,q3,q4,q5,q6,q9}. The representative valley materials are transition-metal
dichalcogenide (TMD) monolayers with the missing  inversion symmetry\cite{q6}, where   two  well separated inequivalent valleys can exist, and
their energies are degenerate within spin-orbit coupling (SOC) interaction. To take advantage of the valley degree of freedom,
the valley polarization should be induced, and several strategies have been proposed, such as  the optical pumping, magnetic field, magnetic
substrates and  magnetic doping\cite{q9-1,q9-2,q9-3,q9-4}. However,  these methods limits the
valleytronics developments due to very weak valley polarization.  The time-reversal symmetry  of two-dimensional (2D) magnetic semiconductors is naturally broken, which  can give rise to spontaneous valley polarization together with the SOC interactions, namely FV
material\cite{q10}.  Many FV  materials have been predicted by the density functional theory (DFT) calculations\cite{v1,v2,v3,v4,v5,v6,q11,q12,q13,q13-1,q14,q15,q16,q17}. Recently, a new concept of HVM has been proposed\cite{v4}, being analogous to half metals
in spintronics. The one valley of HVM is metallic,  while the other
is semiconducting, which possesses  100\% valley polarization.

Since  the quantum spin Hall (QSH) state is discovered  in
graphene\cite{t1,t2}, the QSH  materials have attracted tremendous attention.
The QSH insulator  has a bulk energy gap with topologically protected helical edge states,
which  are protected by topology and are robust against backscattering.
The breaking time-reversal symmetry  of 2D topological
insulator results in the QAH effect, which  can display quantized
Hall conductance under zero magnetic field.  Experimentally, the QAH insulator is firstly observed
 in Cr doped
$\mathrm{(Bi, Sb)_2Te}$ thin films  below 30 mK\cite{qa1}. A higher-temperature
QAH insulator is still a  challenge in experiment\cite{qa1,qa2}.
Recently, some robust QAH insulator are proposed with
high Curie temperature\cite{fe,fe1,fe2}.

\begin{figure*}
  \includegraphics[width=11cm]{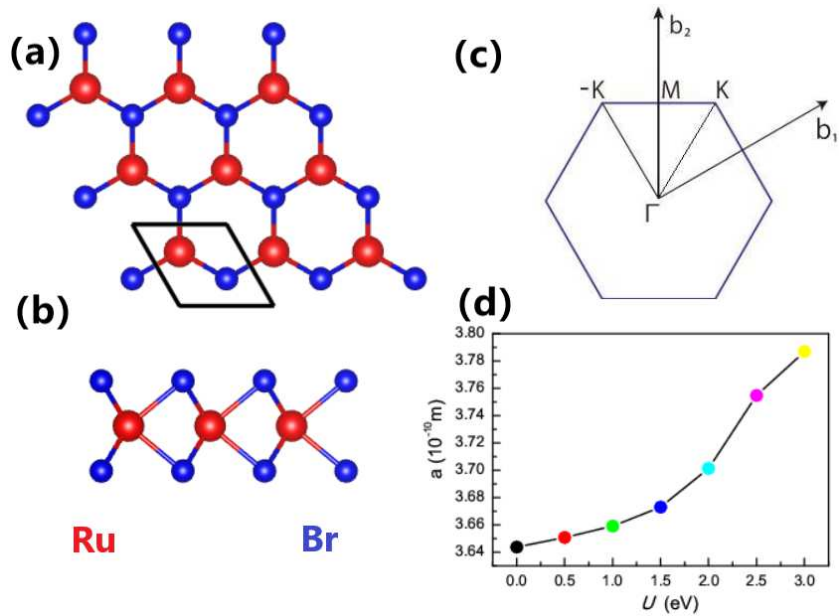}
  \caption{(Color online)For  $\mathrm{RuBr_2}$ monolayer, the (a) top view and (b) side view of  crystal structure,  and (c) the Brillouin zone with high-symmetry points labeled, and  (d) the lattice constants $a$ as a function of $U$.}\label{st}
\end{figure*}
\begin{figure}
  \includegraphics[width=7cm]{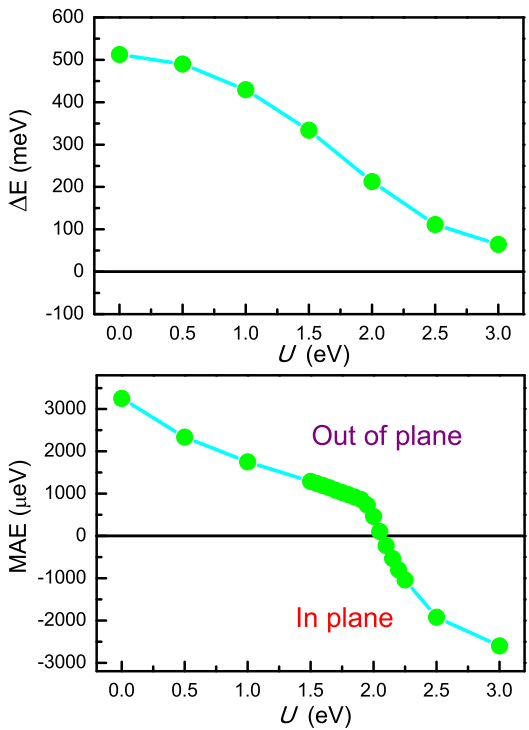}
  \caption{(Color online)For $\mathrm{RuBr_2}$ monolayer, the  energy differences $\Delta E$ with rectangle supercell between  AFM and FM ordering  and MAE as a function of $U$.}\label{u-em}
\end{figure}

It's a natural idea to combine  valleytronics and band topology,  namely VQAHI.  The valley-polarized quantum anomalous Hall effect (VQAHE)  can be realized  in the Co decorated In-triangle adlayer on a Si(111) surface\cite{t3}, metal-organic frameworks (MOFs)\cite{t4}, bilayer graphene with
layer-dependent proximity effects\cite{t5},  MXenes\cite{t6}, bilayer graphene subject  interlayer bias and light irradiation\cite{t7}.
However, it is highly desirable to search for a VQAHI, which  is experimentally feasible in 2D materials with simple lattice structure.
Recently, the $\mathrm{RuBr_2}$ monolayer with $1H$-$\mathrm{MoS_2}$ type structure is predicted to become VQAHI by strain\cite{t8}. However, the electronic correlation and magnetic anisotropy
are not detailedly taken into account during the calculations of electronic structures, which are very key to produce novel electronic states\cite{v4,t9,t10}.
In this work, the electronic correlation and magnetic anisotropy  on  electronic structures of $\mathrm{RuBr_2}$ monolayer are comprehensively investigated by using  GGA+$U$+SOC approach.  It is found
that different  correlation strengths ($U$) along with different magnetic anisotropy (out-of-plane and  in-plane) can induce
different electronic states.  For PMA,  the increasing $U$ can drive the system into FV to HVM  to QAH to HVM to FV states. But,
 there are no novel QAH states and observable valley polarization for in-plane situation.
In the appropriate $U$ range,  the $\mathrm{RuBr_2}$ monolayer indeed possesses PMA, QAH states and spontaneous valley polarization, which is a VQAHI.
Even though the real $U$  is beyond the scope of achieving  VQAHI, it can be realized by strain.  Our works highlight the importance  of
correlation effects and magnetic anisotropy in $\mathrm{RuBr_2}$ monolayer.

The rest of the paper is organized as follows. In the next
section, we shall give our computational details and methods.
 In  the next few sections,  we shall present structure and magnetic properties, electronic correlation and strain effects on  electronic properties and Curie temperature of  $\mathrm{RuBr_2}$ monolayer. Finally, we shall give our discussion and conclusion.

\begin{figure}
  \includegraphics[width=7cm]{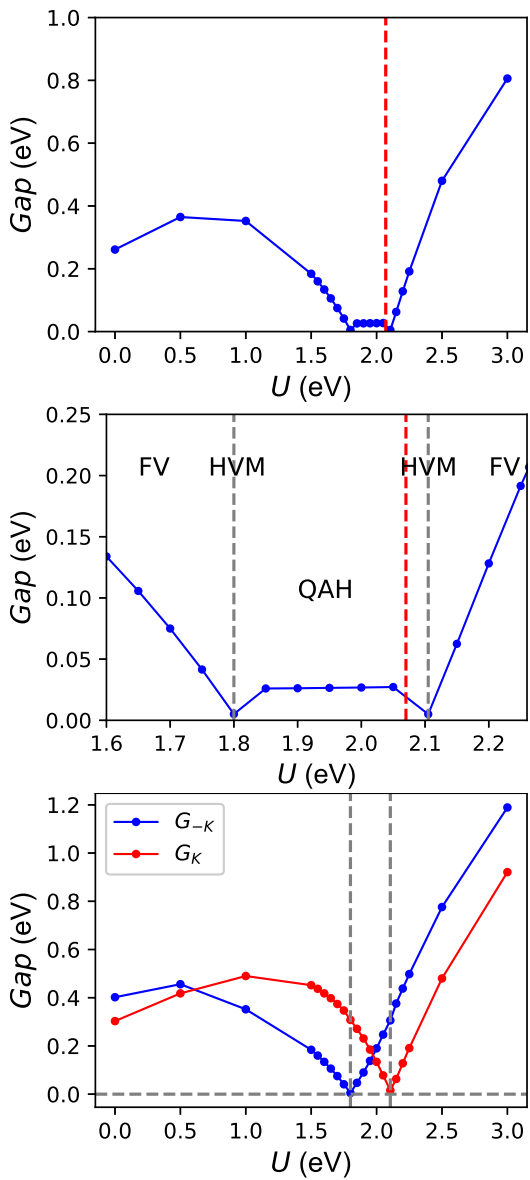}
  \caption{(Color online)For $\mathrm{RuBr_2}$ monolayer with out-of-plane  magnetic anisotropy, the top plane shows  global energy band gap  as a function of   $U$ (0-3 eV). The middle  plane is the enlarged view of the top plane  between  $U$$=$1.60 eV and 2.25 eV, and   the phase diagram  is shown with different $U$ region. The bottom plane  shows  the energy  band gaps for the -K and K valleys. The vertical red dotted line means that actual MAE changes from out-of-plane to in-plane. }\label{u-gap}
\end{figure}

\begin{figure*}
  \includegraphics[width=15cm]{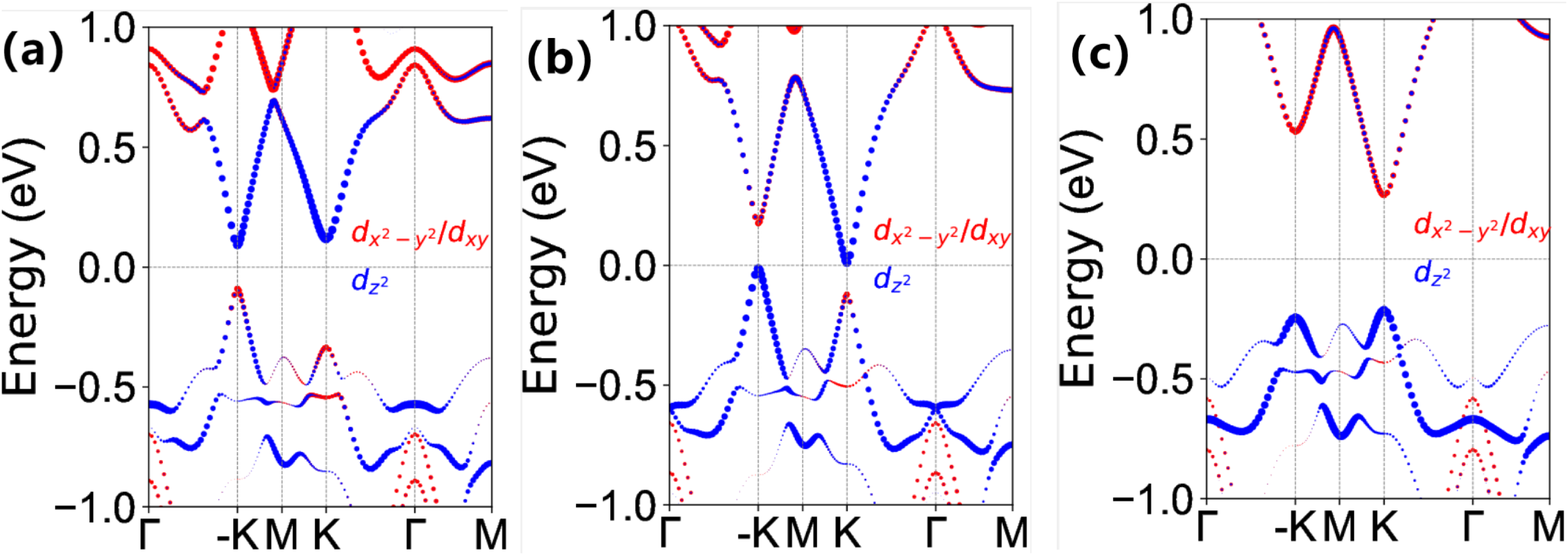}
\caption{(Color online)For out-of-plane magnetic anisotropy, the Ru-$d_{x^2-y^2}$/$d_{xy}$ and $d_{z^2}$-orbital characters of energy bands of  monolayer $\mathrm{RuBr_2}$ with  $U=$1.5 eV (a), 2.0 eV (b) and 2.5 eV (c).}\label{d-s}
\end{figure*}

\section{Computational detail}
We perform  spin-polarized  first-principles calculations    within DFT\cite{1},  as implemented in VASP code\cite{pv1,pv2,pv3}. The projected
augmented wave (PAW) method is adopted, and
  the GGA of Perdew-Burke-Ernzerhof (PBE-GGA)\cite{pbe} is used as exchange-correlation potential.
The energy cut-off of 500 eV is used to attain accurate results. The total energy  convergence criterion of  $10^{-8}$ eV  is used, and
the force
convergence criteria of  being  less than 0.0001 $\mathrm{eV.{\AA}^{-1}}$ is set on each atom.  A vacuum space  with a thickness of more than 18 $\mathrm{{\AA}}$ is used  to ensure decoupling between the periodic
layers. The $\Gamma$-centered 24$\times$24$\times$1 k-mesh is  employed to sample the Brillouin zone for structure optimization, electronic structures and elastic properties, and 12$\times$24$\times$1 Monkhorst-Pack k-point mesh for FM and antiferromagnetic (AFM)  energies with rectangle supercell.
The   GGA+$U$  method within  the
rotationally invariant approach proposed by Dudarev et al is employed to describe the strong correlated  Ru-$d$ electrons, where only the effective
$U$ ($U_{eff}$) by subtracting exchange parameters  from the on-site Coulomb interaction
parameter  is meaningful.
The SOC effect is explicitly included to investigate MAE, electronic and topological properties of $\mathrm{RuBr_2}$ monolayer.

We use strain-stress relationship (SSR) method to attain  elastic stiffness tensor  $C_{ij}$.
The  2D elastic coefficients $C^{2D}_{ij}$ have been renormalized by   $C^{2D}_{ij}$=$L_z$$C^{3D}_{ij}$, where  the $L_z$  is  the length of unit cell along z direction. The  edge states  are calculated with the maximal localized
Wannier function tight-binding model by employing the $d$-orbitals of Ru atoms and the  $p$-orbitals of Br atoms\cite{w1,w2}.
The Berry curvatures of $\mathrm{RuBr_2}$ monolayer
are calculated directly from the calculated
wave functions  based on Fukui's
method\cite{bm} by using the VASPBERRY code\cite{bm1,bm2}.
We use  40$\times$40 supercell and  $10^7$ loops to perform  the
Monte Carlo (MC) simulations for calculating Curie temperature $T_C$ of  $\mathrm{RuBr_2}$ monolayer, as implemented in Mcsolver code\cite{mc}.

\begin{figure*}
  \includegraphics[width=12cm]{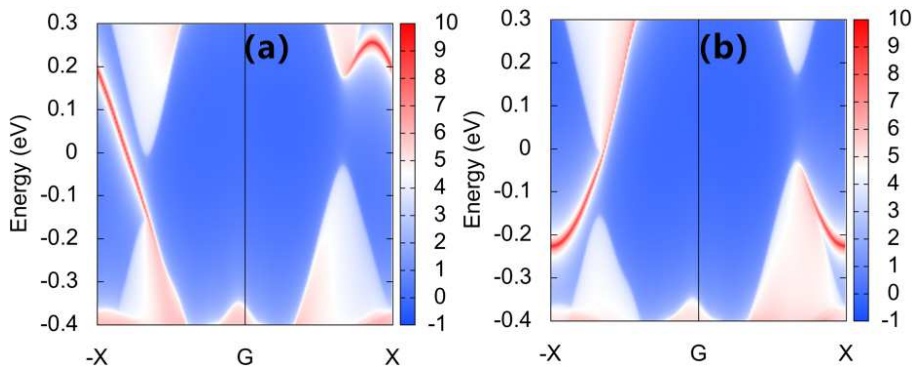}
\caption{(Color online)For $\mathrm{RuBr_2}$ monolayer with out-of-plane  magnetic anisotropy,  the topological left (a) and right (b)
edge states  calculated  along the (100) direction with $U$ being 2.00 eV.}\label{u-s}
\end{figure*}

\begin{figure*}
  \includegraphics[width=15cm]{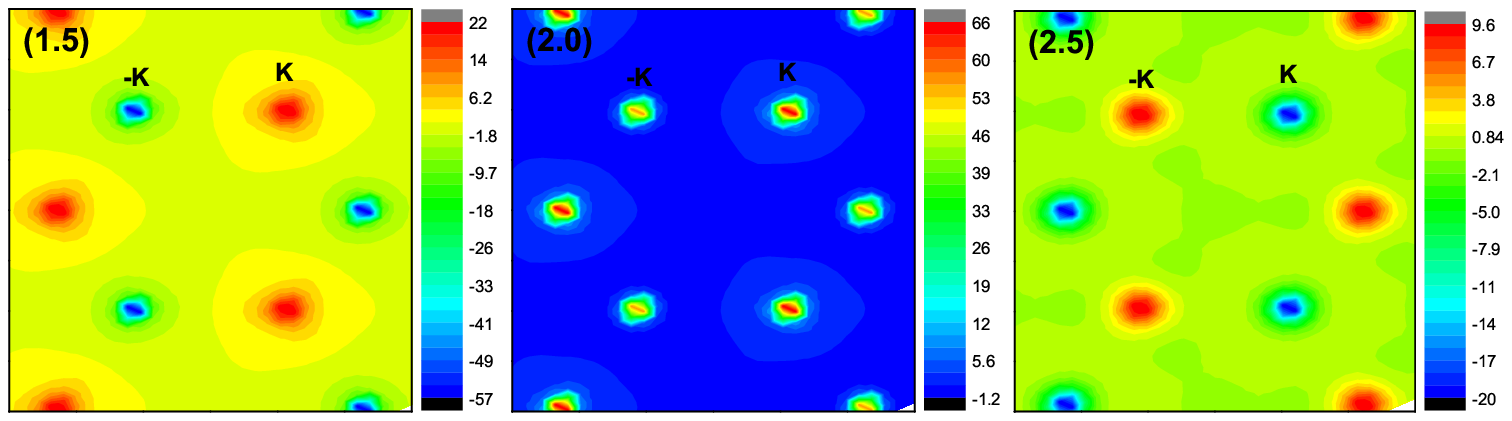}
  \caption{(Color online)For $\mathrm{RuBr_2}$ monolayer with out-of-plane  magnetic anisotropy, the calculated Berry curvature distribution in the 2D Brillouin zone with different $U$ (1.5 eV, 2.0 eV and 2.5 eV).}\label{u-berry}
\end{figure*}

\section{Structure and magnetic properties}
The $\mathrm{RuBr_2}$ monolayer possesses $P\bar{6}m2$  symmetry with space group No.187, which  consists Br-Ru-Br sandwich layer, shown in \autoref{st} along with Brillouin zone with high-symmetry points. The $\mathrm{RuBr_2}$ monolayer shares the same crystal structure with classical 2D material 1$H$-$\mathrm{MoS_2}$. Its inversion symmetry  is broken, which can lead to many novel properties, such as piezoelectricity and valley features. Different from the previous works\cite{v4,t8,t9}, we optimize lattice constants $a$ of $\mathrm{RuBr_2}$ monolayer at different $U$ (0-3 eV), as illustrated
in \autoref{st}. With increasing $U$,  the $a$ increases from  3.644 $\mathrm{{\AA}}$ to 3.787 $\mathrm{{\AA}}$.
To obtain the magnetic ground states, we calculate the
total energy difference between  AFM and FM ordering by using rectangle supercell as a function of $U$, as plotted in \autoref{u-em}.  The
FM states are found to be the magnetic ground states for all considered $U$ values, and the increasing $U$
  weaken FM interaction, which has important influence on $T_C$ of  $\mathrm{RuBr_2}$ monolayer.
The direction of magnetic anisotropy is very critical to determine topological and valley properties of some 2D materials\cite{t9,t10}.
Here, the  MAE is used to determine  intrinsic magnetic anisotropy of $\mathrm{RuBr_2}$ monolayer at different $U$ value by calculating $E_{MAE}$ = $E_{(100)}$-$E_{(001)}$ within GGA+SOC+$U$, which is shown in  \autoref{u-em}. The positive $E_{MAE}$ means out-of-plane magnetic anisotropy, while the negative value suggests the in-plane one. With increasing $U$, the magnetic anisotropy direction changes  from out-of-plane to in-plane one, and the critical $U$ value is about 2.07 eV

\begin{figure}
   \includegraphics[width=8cm]{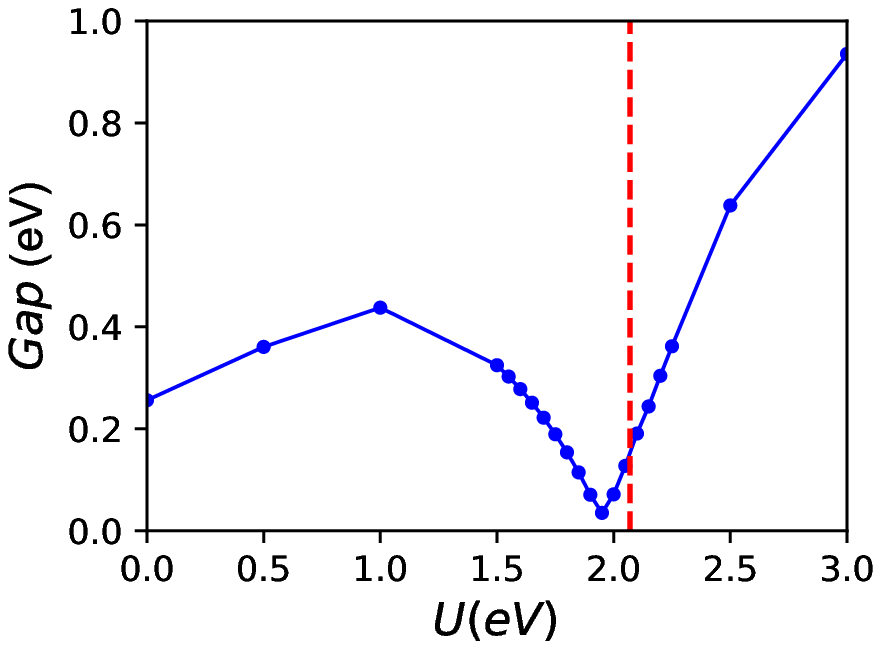}
  \caption{(Color online)For $\mathrm{RuBr_2}$ monolayer with in-plane magnetic anisotropy, the energy band gap as a function of   $U$ (0-3 eV). The vertical red dotted line means that actual MAE changes from out-of-plane to in-plane.}\label{u-gap-1}
\end{figure}

\begin{figure}
  \includegraphics[width=7cm]{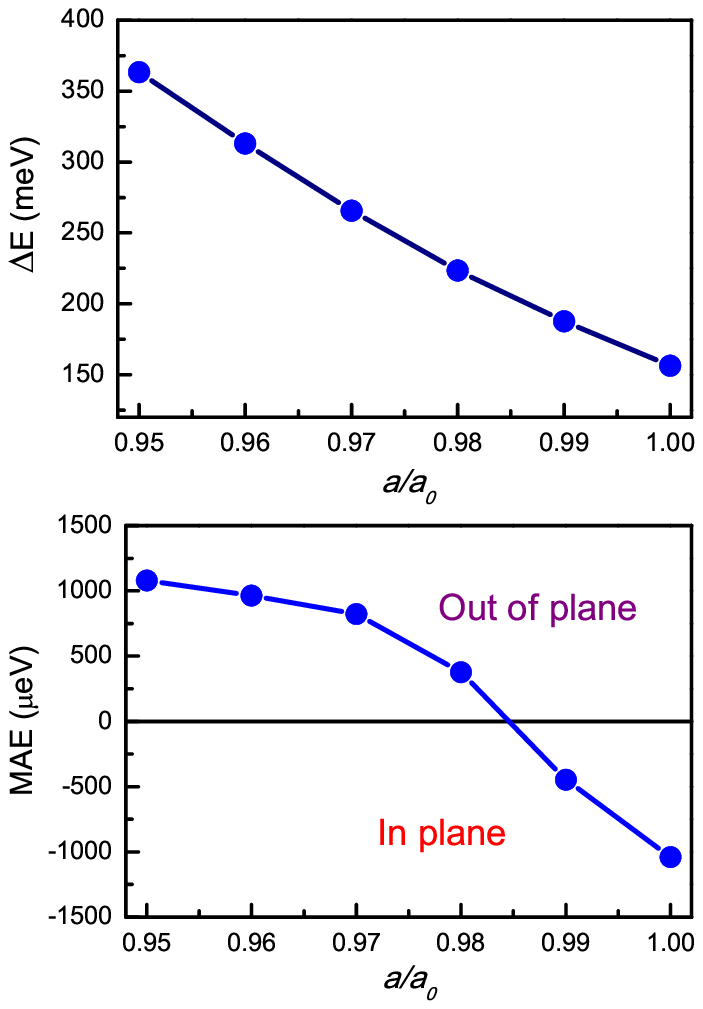}
  \caption{(Color online)For $\mathrm{RuBr_2}$ monolayer, the  energy differences $\Delta E$ with rectangle supercell between  AFM and FM ordering  and MAE as a function of strain $a/a_0$.}\label{s-em}
\end{figure}
\begin{figure}
  \includegraphics[width=7cm]{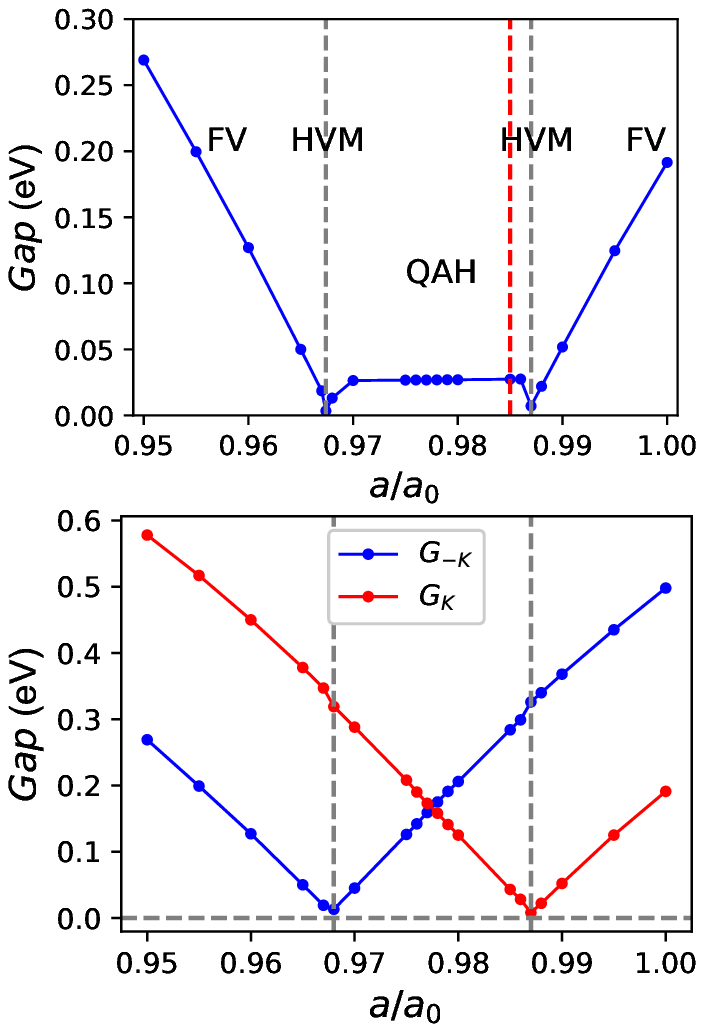}
  \caption{(Color online)For $\mathrm{RuBr_2}$ monolayer with out-of-plane  magnetic anisotropy, the top plane shows  global energy band gap  as a function of  strain $a/a_0$ (0.95-1.00), and   the phase diagram  is shown with different strain region. The bottom plane  shows  the energy  band gaps for the -K and K valleys. The vertical red dotted line means that actual MAE changes from out-of-plane to in-plane.}\label{s-gap}
\end{figure}

The dynamical and thermal stabilities of $\mathrm{RuBr_2}$ monolayer have been proved by phonon spectra without soft modes and the
molecular-dynamics simulations\cite{t8}. To further check its
mechanical stability,  the elastic
properties of  $\mathrm{RuBr_2}$ monolayer are investigated.
Due to $P\bar{6}m2$  space group,   there are two independent elastic constants: $C_{11}$ and $C_{12}$, and the corresponding values are
 40.29 $\mathrm{Nm^{-1}}$ and 14.57 $\mathrm{Nm^{-1}}$ by using GGA.  The calculated elastic constants meet the  Born  criteria of mechanical stability\cite{ela}: $C_{11}$$>$0 and $C_{11}-C_{12}$$>$0,    indicating  its  mechanical stability.
  Due to hexagonal symmetry, the  Young¡¯s moduli $C^{2D}$, shear modulus $G^{2D}$ and Poisson's ratios $\nu^{2D}$ of $\mathrm{RuBr_2}$ monolayer all  are mechanically isotropic, and the calculated results show that they  are 35.02 $\mathrm{Nm^{-1}}$, 12.86 $\mathrm{Nm^{-1}}$ and 0.362, respectively.

\begin{figure*}
  \includegraphics[width=12cm]{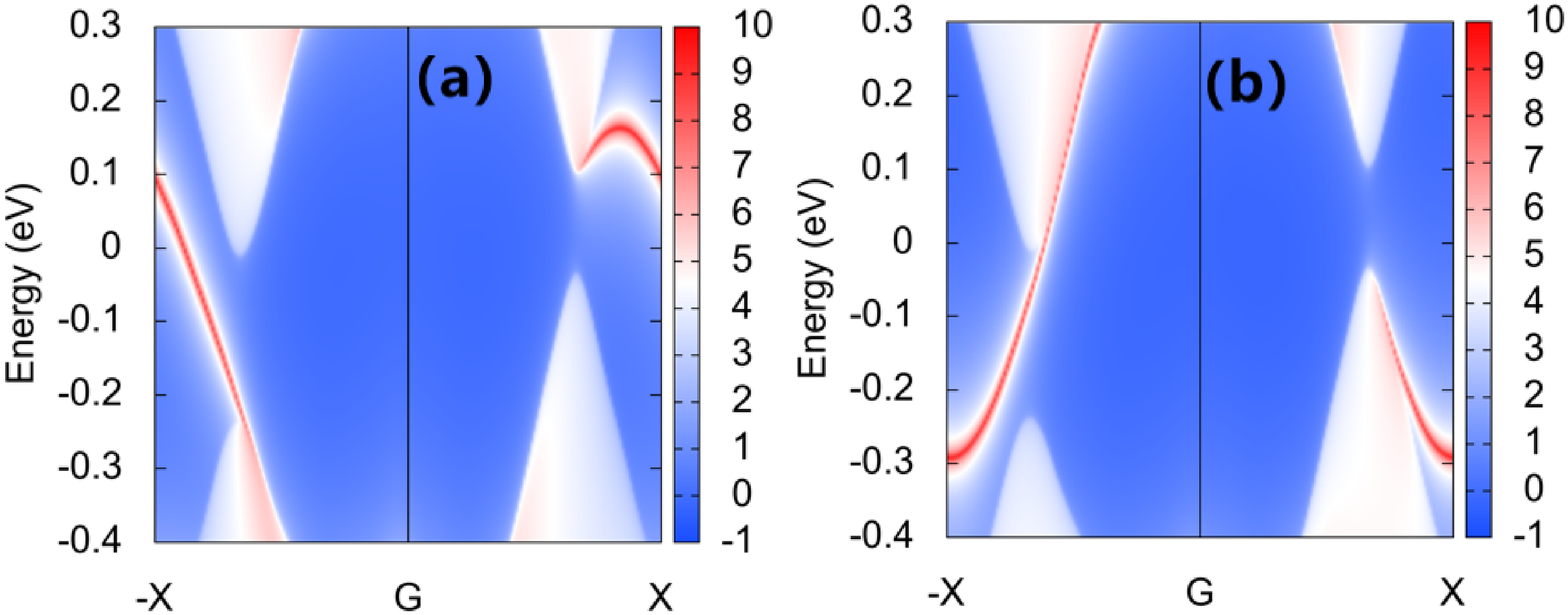}
\caption{(Color online)For $\mathrm{RuBr_2}$ monolayer with out-of-plane  magnetic anisotropy,  the topological left (a) and right (b)
edge states  calculated  along the (100) direction with $a/a_0$ being 0.975.}\label{s-s}
\end{figure*}

\section{Electronic correlation effects}
Electronic correlations  have significant impact on the magnetic, topological and electronic properties of 2D materials\cite{v4,t9,t10}. The direction of magnetic anisotropy can  affect the symmetry of 2D materials,  and then produce important influence on
the electronic and topological properties\cite{n1,t9,t10}. Firstly, the magnetocrystalline direction  of
$\mathrm{RuBr_2}$ monolayer is assumed to be along the out-of-plane one in considered $U$ range (0-3 eV).
The out-of-plane magnetic anisotropy  allows a nonvanishing Chern
number of 2D materials with varied $U$\cite{n1}.
The electronic structures of $\mathrm{RuBr_2}$ monolayer with different $U$ are calculated by GGA+SOC+$U$. The energy band structures at some representative $U$
values are plotted in FIG.1 of electronic supplementary information (ESI), and the evolutions of energy band gaps  vs $U$ are plotted in \autoref{u-gap}, along with the gaps at -K and K points.

For    $U$$<$1.8 eV, the gap firstly increases, and then decreases, which is due to the transformation of valence band maximum (VBM) .
At small $U$, the $\mathrm{RuBr_2}$ monolayer is an indirect
gap semiconductor. The conduction band minimum
(CBM) is at
the -K point, whereas the VBM  occurs at the one point being close to M point. With increasing $U$,  the system becomes a direct
gap semiconductor with both VBM and CBM being at -K point. Around the  $U$$=$1.8 eV,  the HVM with conduction electrons being intrinsically 100\% valley polarized\cite{v4} can be achieved, whose gap at -K point  gets closed, and a band gap of 0.31 eV still holds at K point.
 When the $U$  reaches  2.105 eV,  the other HVM
state can be observed, where the -K valley holds the  energy  band
gap (0.31 eV) and the K valley is metallic,  opposite to the  previous case.
Between $U$=1.8 eV and 2.105 eV, the gap of about 26 meV can be observed, which is  larger than
the thermal energy of  room temperature (25 meV).   However, for  $U$$>$2.105 eV,  the gap increases  with increasing $U$, and the $\mathrm{RuBr_2}$ monolayer changes from a direct
gap semiconductor to an indirect
gap semiconductor with CBM always being at K point.

In all regions, $\mathrm{RuBr_2}$ monolayer is a FV material. The valley polarization can be observed  in both valence and condition bands at the -K and K points, and the corresponding valley splitting is plotted in FIG.2 of ESI.
For $U$$<$1.8 eV, a remarkable valley splitting  appears  in the  valence bands. However, for $U$$>$2.105 eV,  the noteworthy valley polarization can be observed in the conduction bands.  Between $U$$=$1.8 eV and 2.105 eV, with increasing $U$, the valley splitting in the valence bands decreases, and increases for the conduction bands. These can be understood by the transformation of  Ru-$d$-orbital characters of energy bands. In all $U$ region, the  $d_{z^2}$ or $d_{x^2-y^2}$/$d_{xy}$ orbitals of Ru atoms dominate  the -K and K valleys of both valence and conduction bands. At representative $U$ (1.5 eV, 2.0 eV and 2.5 eV), the Ru-$d_{x^2-y^2}$/$d_{xy}$ and $d_{z^2}$-orbital characters of energy bands of  monolayer $\mathrm{RuBr_2}$ are plotted in \autoref{d-s}.
For  $U$$<$1.8 eV, the $d_{x^2-y^2}$ and $d_{xy}$ orbitals dominate -K and K valleys in valence bands, and the $d_{z^2}$ orbitals play a leading role for those in the conduction bands (see \autoref{d-s} (a)).
For  $U$$>$2.105 eV, the  situation is opposite to the  previous case of  $U$$<$1.8 eV for the distribution of Ru-$d$-orbitals (see \autoref{d-s} (c)).

The SOC induces  valley polarization, which is contributed by the intra-atomic interaction:
\begin{equation}\label{m1}
\hat{H}_{SOC}=\lambda\hat{L}\cdot\hat{S}=\hat{H}^0_{SOC}+\hat{H}^1_{SOC}
\end{equation}
where  $\lambda$, $\hat{L}$ and $\hat{S}$ are the coupling strength, the orbital angular moment and spin angular moment, respectively.
Due to the magnetic exchange interaction,
 the $\hat{H}^0_{SOC}$  dominates the  $\lambda\hat{L}\cdot\hat{S}$.
For out-of-plane magnetization, the  $\hat{H}^0_{SOC}$  can be expressed as:
\begin{equation}\label{m1}
\hat{H}^0_{SOC}=\alpha \hat{L}_z
\end{equation}
The orbital
basis for -K and K valleys  with the group symmetry being $C_{3h}$  can be written as\cite{q10,v2,v3}:
\begin{equation}\label{m2}
   \begin{array}{c}
|\phi^\tau>=\sqrt{\frac{1}{2}}(|d_{x^2-y^2}>+i\tau|d_{xy}>)\\
or\\
|\phi^\tau>=|d_{z^2}>
  \end{array}
\end{equation}
in which  the subscript  $\tau$ represent  valley index ($\tau=\pm1$). At K
and -K valleys,  the resulting energy   can be expressed as:
\begin{equation}\label{m3}
E^\tau=<\phi^\tau|\hat{H}^0_{SOC}|\phi^\tau>
\end{equation}
If the -K and K valleys are dominated by $d_{x^2-y^2}$ and $d_{xy}$ orbitals, the valley splitting $|\Delta E|$
  can be expressed as:
\begin{equation}\label{m4}
|\Delta E|=E^{K}-E^{-K}=4\alpha
\end{equation}
For the $d_{z^2}$ orbitals, the valley splitting $|\Delta E|$
 is given as:
\begin{equation}\label{m4}
|\Delta E|=E^{K}-E^{-K}=0
\end{equation}
If considering $\hat{H}^1_{SOC}$, the valley splitting $|\Delta E|$ at K
and -K valleys
mainly from the $d_{z^2}$ orbitals is not equal to zero. These can explain $U$ dependence of valley splitting for conduction or valence bands.

 When $U$ is between  1.8 eV and 2.105 eV, the $d_{x^2-y^2}$ and $d_{xy}$ orbitals dominate  the -K valley in the conduction bands and K valley in the valence band, and the $d_{z^2}$ orbitals dominate  -K valley in the valence  bands and K valley in the conduction band (see \autoref{d-s} (b)). This is a transitional region
 to achieve  the changes of  distribution of Ru-$d$-orbitals with  $U$$<$1.8 eV  to  $U$$>$2.105 eV. In other words, it can be seen that
the band inversion of $d_{xy}$/$d_{x^2-y^2}$ and $d_{z^2}$ orbitals  occurs at -K valley with $U$ across 1.8 eV, and the  band inversion appears again with $U$  crossing 2.105 eV.
 It is noted that the gap of $\mathrm{RuBr_2}$ monolayer closes, reopens, and then closes. The band inversion and novel change of gap   suggest some topological phase transitions, and  QAH insulator phases may exist  between the two HVM states. Similar phenomenon can be observed in $\mathrm{FeCl_2}$ and FeClF monolayers\cite{v4,t10}.
The chiral edge states are calculated to confirm  QAH  phases, which are plotted in \autoref{u-s}  with $U$$=$2.00 eV.
A chiral edge state  does exist for  left/right edge, which connects the conduction bands and
valence  bands.  The Chern number is equal to one ($C$=1) due to a single  chiral edge state. The $C$=1 can also be obtained
by integrating the Berry curvature (see \autoref{u-berry}) within the first Brillouin zone
of $\mathrm{RuBr_2}$ monolayer, which is consistent with a single gapless chiral edge band.
It is clearly seen that a valley structure for both conduction and valence bands can coexist  with QAH phase (See FIG.1 of ESI), and the K/-K valley can be polarized with the Fermi level  into conduction/valence bands.  That is  also called VQAHI, which can combine valleytronics and spintronics with nontrivial band topology. With increasing $U$, the $\mathrm{RuBr_2}$ monolayer  can undergo  the FV, HVM, QAH, HVM and FV states.

These electronic state transitions are related with the variation of Berry curvature. In considered $U$ range, the Berry curvature is calculated, which is shown in \autoref{u-berry} at representative $U$
values ($U$=1.5, 2.0 and 2.5 eV). In all considered  $U$ range, the
hot spots of  Berry curvature  are around -K and K valleys .  For  $U$$<$1.8 eV and  $U$$>$2.105 eV, the
 Berry curvatures  around two valleys have opposite
signs and different magnitudes. However, for 1.8 eV$<$$U$$<$2.105 eV, they have the same signs.
   With increasing $U$,  a topological phase transition is produced, which is  connected by the first  HVM state.  In this transition, the sign of Berry curvature  at -K valley flips. For example, the negative
  Berry curvature (-K valley) at  $U$$=$1.5 eV  changes into positive one at $U$$=$2.0 eV. When $U$ continues to increase (The $U$ spans the 2.105 eV), another  topological phase transition induces  the sign flipping  of Berry curvature  at K valley. For instance, the positive  Berry curvature of K valley at $U$$=$2.0 eV changes into negative one at $U$$=$2.5 eV. These mean that the electronic correlation effects can induce sign-reversible  Berry curvature, which is related with topological phase transition.

Next, we assume that the magnetocrystalline direction  of
$\mathrm{RuBr_2}$ monolayer is in-plane.
By using GGA+SOC, the energy band gaps as a function of  $U$ are plotted  in \autoref{u-gap-1}, and the representative energy band structures  are shown  in FIG.3 of ESI.  With increasing $U$, the gap firstly decreases, and  reduces to zero (about $U$$=$1.95 eV), and increases again.
It is clearly seen that no QAH region appears, and no observable valley polarization  in both the valence and conduction bands exists.
Calculated results show that the distributions of Ru-$d$-orbital characters are akin to the cases of out-of-plane, and the sign-reversible Berry curvature can also be observed, when $U$ increases across about 1.95 eV. The
hot spots of  Berry curvature  around two valleys  have opposite
signs and almost the same magnitudes.  According to calculated results in \autoref{u-em}, the magnetocrystalline direction  of $\mathrm{RuBr_2}$ monolayer is out-of-plane with $U$ being less than 2.07 eV, and is in-plane with $U$ being greater than 2.07 eV.  For  $U$$<$2.07 eV, there can be FV, HVM and QAH states. However, only common FM semiconductors can exist with  $U$$>$2.07 eV. As long as the actual $U$ is between 1.8 eV and 2.07 eV, the $\mathrm{RuBr_2}$ monolayer intrinsically is a VQAHI.

\begin{figure*}
  \includegraphics[width=15cm]{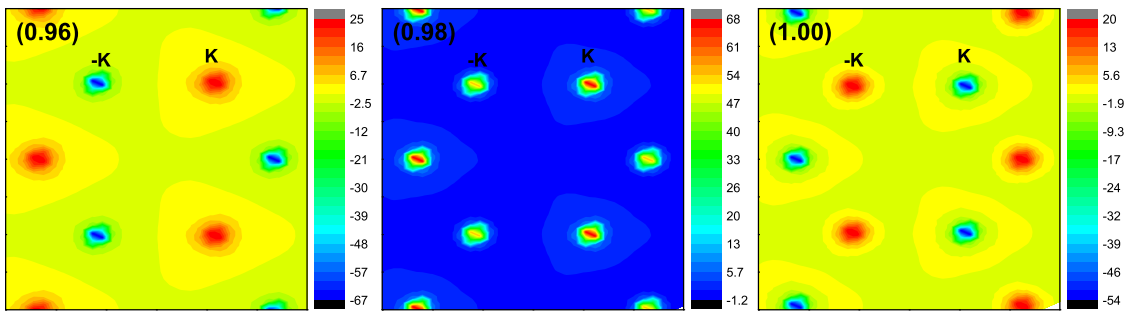}
  \caption{(Color online)For $\mathrm{RuBr_2}$ monolayer with out-of-plane  magnetic anisotropy, the calculated Berry curvature distribution in the 2D Brillouin zone with different $a/a_0$ (0.96, 0.98 and 1.00).}\label{s-berry}
\end{figure*}

\begin{figure}
   \includegraphics[width=8cm]{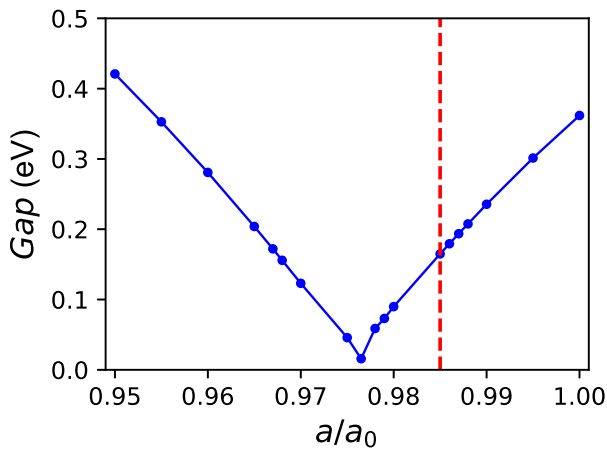}
  \caption{(Color online)For $\mathrm{RuBr_2}$ monolayer with in-plane magnetic anisotropy, the energy band gap as a function of strain  $a/a_0$ (0.95-1.00). The vertical red dotted line means that actual MAE changes from out-of-plane to in-plane.}\label{s-gap-1}
\end{figure}

\section{Strain effects}
 For a given material, the correlation strength $U$ should be fixed, and should
be determined from experiment.  The competition between kinetic and interaction energies determines the strength of electronic correlation\cite{t9}. The strain can modify the distance between atoms, which can tune kinetic and interaction energies. For hexagonal lattice with a
formula of $\mathrm{MX_2}$,   by means of a low energy $k\cdot p$ model analysis, a
mechanism of producing VQAHI  via strain engineering is
proposed, and a general picture of valley-contrasted band inversion is developed\cite{v7}. Then the effective Hamiltonian from strain contribution can be written as\cite{v7}:
\begin{equation}\label{pe0-1-1}
H_\varepsilon=\frac{D(\varepsilon_{xx}+\varepsilon_{yy})\hat{s_0}\hat{\sigma_z}}{2}=\frac{D\varepsilon_{xx}\hat{s_0}\hat{\sigma_z}}{2}
\end{equation}
where $D$ and $\varepsilon_{xx}$/$\varepsilon_{yy}$ are deformation potential and strain along x/y axis, $\hat{s_0}$ and $\hat{\sigma_z}$ denote the Pauli matrices. The $H_\varepsilon$ can induce  band inversion, producing QAH phase.

Therefore, the $\mathrm{RuBr_2}$ monolayer can be tuned into FV, HVM and QAH states, even though the actual $U$ is larger than 2.07 eV. We use $U$$=$2.25 eV as a example to  prove the assumption, where the $\mathrm{RuBr_2}$ monolayer is a common FM semiconductor.  The $a/a_0$ is used  to simulate the biaxial strain, in which  $a$ and $a_0$ are the strained and  unstrained lattice constants. Here, the compressive strain ($a/a_0$$<$1) is applied to achieve the transition of electronic states.
To confirm the FM ground states, the
total energy differences between  AFM and FM ordering by using rectangle supercell are calculated  as a function of $a/a_0$, as shown in \autoref{s-em}. In considered strain range,  the FM states are always the magnetic ground states, and the compressive strain
  can enhance  FM interaction,  improving  $T_C$ of  $\mathrm{RuBr_2}$ monolayer.

Firstly, we assume that the magnetocrystalline direction  of
$\mathrm{RuBr_2}$ monolayer is  along the out-of-plane one. At some representative $a/a_0$,
the energy band structures  are plotted in FIG.4 of ESI, and the energy band gaps along with the gaps at -K and K valleys as a function of $a/a_0$ are shown in \autoref{s-gap}.
For    $a/a_0$$>$0.987, the gap  decreases with increasing compressive strain.
Around the  $a/a_0$$=$0.987,  the  first HVM  can be achieved, whose gap at K point  gets closed.
 When the $a/a_0$  reduces to  0.968,  the other HVM
state can be observed, where  the -K valley is metallic,  opposite to the  previous case.
Between $a/a_0$=0.987 and 0.968, there is a gap of about 27 meV, which is close to
the thermal energy of  room temperature.
For  $a/a_0$$<$0.968,  the gap increases  with increasing compressive strain.
In considered  strain range, $\mathrm{RuBr_2}$ monolayer is always  a FV material, which can be confirmed by valley splitting in FIG.5 of ESI.
For $a/a_0$$<$0.968, a remarkable valley splitting  can be observed   in the  valence bands, while the noteworthy valley polarization exists in the conduction bands for  $a/a_0$$>$0.987.
Between $a/a_0$$=$0.968 and 0.987, with increasing compressive strain, the valley splitting in the valence bands increases, and decreases for the conduction bands. These can also  be explained by the transformation of  Ru-$d$-orbital characters of energy bands with variational $a/a_0$ , being analogous to changed $U$.

 With increasing compressive strain, the gap of $\mathrm{RuBr_2}$ monolayer  gets closed, reopens, and then closes, which  suggests that  QAH insulator phases may exist  between the two HVM states.
 To confirm  QAH  phases, the chiral edge states  are plotted in \autoref{s-s}  at $a/a_0$$=$0.975.
It is clearly seen that a chiral edge state  for  left/right edge connects the conduction bands and
valence  bands, which means that the Chern number is equal to one ($C$=1).
The  VQAHI can also be observed  between the two HVM states. With increasing compressive strain, the $\mathrm{RuBr_2}$ monolayer  changes from  FV to HVM to QAH to HVM to FV state.  The sign-reversible  Berry curvature can also be observed with increasing compressive strain, which is plotted in \autoref{s-berry} at representative $a/a_0$
values. Similar phenomenon can be found  in  monolayer $\mathrm{VSi_2N_4}$\cite{v8}. With reduced $a/a_0$, the  sign of Berry curvature  at K valley firstly  flips at about $a/a_0$=0.987, and then the Berry curvature  at -K valley change sign at about $a/a_0$$=$0.968. These topological phase transitions are also related with the band inversion of $d_{xy}$/$d_{x^2-y^2}$ and $d_{z^2}$ orbitals, being  similar to varied $U$.

Next,  the magnetocrystalline direction  of
$\mathrm{RuBr_2}$ monolayer is  supposed to be in-plane.
The energy band gaps vs  $a/a_0$ are plotted  in \autoref{s-gap-1}, and the representative energy band structures  are plotted  in FIG.6 of ESI.
With increasing compressive strain,  the gap firstly decreases, and  then increases. The corresponding critical $a/a_0$ is about 0.976.
There is not  QAH phases and  observable valley polarization  in both the valence and conduction bands.
The  MAE vs $a/a_0$ is calculate  to determine  intrinsic magnetic anisotropy of $\mathrm{RuBr_2}$ monolayer, which is plotted in  \autoref{s-em}.
 Based on calculated results, the magnetocrystalline direction  of $\mathrm{RuBr_2}$ monolayer is in plane  with $a/a_0$ being larger  than 0.985, and is out-of-plane with $a/a_0$ being less than 0.985.  For  $a/a_0$$<$0.985, there can be QAH, HVM and FV states. However, for $a/a_0$$>$0.985, there is only common FM semiconductor. The VQAHI can indeed be achieved by strain, which possesses intrinsic PMA, QAH state and spontaneous valley splitting.

\begin{figure*}
   \includegraphics[width=16cm]{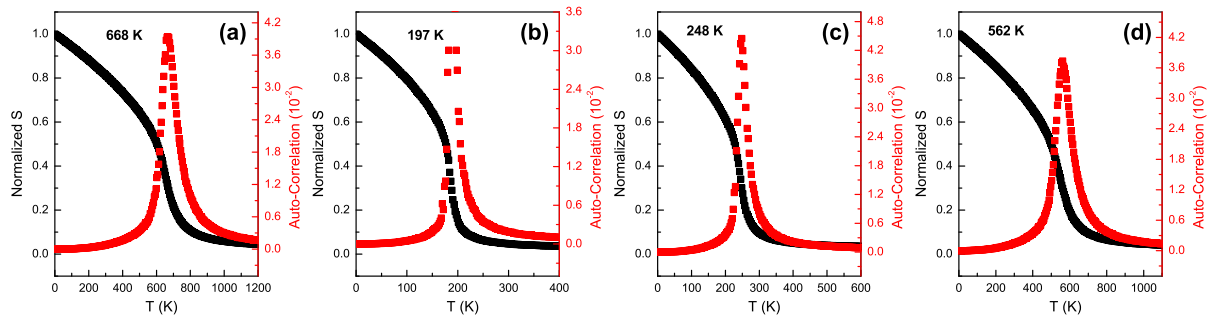}
  \caption{(Color online)For $\mathrm{RuBr_2}$ monolayer, the normalized magnetic moment (S) and auto-correlation  as a function of temperature with $U$$=$1.0 eV (a), 2.5 eV (b) and $a/a_0$=1.0 (c), 0.95 (d) at $U$$=$2.25 eV.}\label{tc}
\end{figure*}
\begin{figure*}
  \includegraphics[width=12cm]{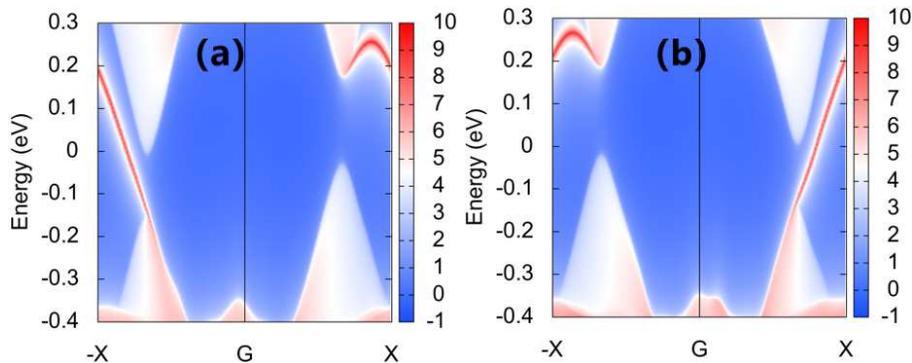}
\caption{(Color online)For $\mathrm{RuBr_2}$ monolayer with out-of-plane  magnetic anisotropy  ($U$$=$2.0 eV),  the topological left
edge states  calculated  along the (100) direction with  SOC for magnetic moment of Ru along the positive (a) and negative (b) z direction, respectively.}\label{s}
\end{figure*}

\section{Curie temperature}
Both electronic correlation effects ($U$) and compressive strain ($a/a_0$) can effectively tune the strength of  FM interaction, which can produce important effects on Curie temperature $T_C$ of monolayer $\mathrm{RuBr_2}$. The $T_C$ is estimated at  representative $U$ (1.0 and 2.5 eV)  and $a/a_0$ (1.0 and 0.95) values
 with the
Wolf   algorithm based on the Heisenberg model, which  can be expressed as:
  \begin{equation}\label{pe0-1-1}
H=-J\sum_{i,j}S_i\cdot S_j-A\sum_i(S_i^z)^2
 \end{equation}
in which   $S_i$/$S_j$ and $S_i^z$   are   the
spin vectors of each Ru atom and  the spin component parallel to the z direction, and $J$ and  $A$ are  the nearest neighbor exchange parameter and   MAE. The $J$ with normalized spin vector ($|S|$=1) can be attained by comparing  energies  of the AFM  ($E_{AFM}$) and FM ($E_{FM}$)
 configurations  with rectangle supercell, and the  corresponding $J$ can be written as:
  \begin{equation}\label{pe0-1-3}
J=\frac{E_{AFM}-E_{FM}}{8}
 \end{equation}
The calculated normalized $J$  are  53.67 meV and 13.95 meV for $U$$=$1.0eV and 2.5 eV, and are 19.55 meV  and 45.41 meV at $a/a_0$$=$1.0 and 0.95 with $U$$=$2.25 eV.
 The   normalized magnetic moment and auto-correlation  vs temperature at  representative $U$ (1.0 and 2.5 eV)  and $a/a_0$ (1.0 and 0.95) values are plotted in \autoref{tc}, and the predicted $T_C$ is about 668/197 K for $U$$=$1.0/2.5 eV, and 248/562 K for $a/a_0$$=$1.0/0.95. So, it is very key for estimating  $T_C$ to use reasonable $U$.  It is found that the compressive strain can improve $T_C$.

\section{Discussion and Conclusion}
The novel electronic state of  $\mathrm{RuBr_2}$ monolayer, like FV, HVM and QAH states, depends on $U$, which  should
be determined from  experiment result.  However, novel electronic state can also be achieved  from common one by strain.
That is to say, the FV, HVM and QAH states can indeed be realized in $\mathrm{RuBr_2}$ monolayer.
These  analysis and results  can be readily extended
to monolayer  $\mathrm{MXY}$ (M = Ru, Os; X/Y=Cl, Br I). The VQAHI
uniquely combines valleytronics and spintronics with nontrivial band topology, which needs appropriate energy band gap and valley splitting  for the practical application. For $\mathrm{RuBr_2}$ monolayer, the band gap of VQAHI is about 26 meV (see \autoref{u-gap} and \autoref{s-gap}), and the valley splitting  can simultaneously be larger than 100 meV by choosing appropriate $U$ or $a/a_0$ (see FIG.2 and FIG.5 of ESI).
These values  are higher than
the thermal energy of  room temperature (25 meV), which makes for  room temperature applications. The VQAHI has  a very special behavior of the chiral-spin-valley  locking for the  edge state. Taking unstrained  $\mathrm{RuBr_2}$ with $U$$=$2.0 eV as a example,  the topological left
edge states   for magnetic moment of Ru along the positive  and negative  z direction are plotted in \autoref{s}.
The valley polarization for bulk can  be switched by reversing the magnetization direction with magnetic moment of Ru along the negative z direction.
Due to the bands all dominated by spin-down bands near the Fermi level, the edge state shown in  \autoref{s} (a) is also spin down
with 100\% spin polarization and 100\% valley polarization. For the -K and K valleys, the edge state only appears at the -K valley, which is  due to the flipping of the sign of the Berry curvature  or band inversion at -K valley. When the magnetization
is reversed, the edge state shown in  \autoref{s} (b) moves  to the K valley,
which has  an opposite spin direction and chiral.

In summary, we have demonstrated the possibility of realizing VQAHI
 in $\mathrm{RuBr_2}$ monolayer, as a representative of the monolayer  $\mathrm{MXY}$ (M = Ru, Os; X/Y=Cl, Br I).
The varied $U$ can tune interplay among  magnetic,
correlation and SOC, which can  result in different electronic state, like  FV, HVM and QAH states.
The VQAHI can be achieved between two HVM states, which possesses  exotic chiral-spin-valley locking edge states.
It is found that strain is a effective method to achieve VQAHI from a common FM semiconductor.
These topological phase transitions  are related with sign-reversible  Berry curvature and band inversions of $d_{xy}$/$d_{x^2-y^2}$ and $d_{z^2}$ orbitals at -K and K valleys. It is found that correlation strength and strain can observably tune magnetic interaction, which is related with Curie temperature. Our works provide a comprehensive understanding  of  $\mathrm{RuBr_2}$ monolayer, which can be used as a candidate material  with complete spin
and valley polarization, combining  valleytronics, spintronics  and topological electronics.

\begin{acknowledgments}
This work is supported by Natural Science Basis Research Plan in Shaanxi Province of China  (2021JM-456) and Graduate Innovation Fund Project in Xi'an University of Posts and Telecommunications (CXJJDL2021001). We are grateful to the Advanced Analysis and Computation Center of China University of Mining and Technology (CUMT) for the award of CPU hours and WIEN2k/VASP software to accomplish this work.
\end{acknowledgments}

\end{document}